   \documentclass{aa}
\usepackage{graphicx}
\begin{document}
   \title{ Spectral Energy Distributions of starburst 
galaxies in the 900-1200 $\rm \AA$ range}

   \author{V. Buat\inst{1}, D. Burgarella
          \inst{1}, J.M. Deharveng\inst{1}
          \and
           D. Kunth\inst{2}}

   \offprints{V. Buat}
   
   \institute{Laboratoire d'Astrophysique de Marseille, BP8, 13376 Marseille 
cedex 12, France\\
        \and
         Institut d'Astrophysique de Paris, Bd Arago, Paris, France
          \\}

   \date{}

   \abstract{ We present the 970-1175 $\rm \AA$ spectral energy 
distributions (SEDs) of 12 starburst galaxies observed with the Far Ultraviolet 
Spectroscopic Explorer {\it FUSE}. 
We take benefit of the high spectral resolution of {\it FUSE} to estimate a 
continuum as much as possible unaffected by 
 the interstellar lines. The continuum is  rather 
flat with, in few cases, a decrease at  $\rm \lambda <~1050 \AA$, the amplitude 
of 
which being correlated with various indicators of the dust extinction.  The 
far-UV SEDs are compared with synthetic population models. The galaxies with 
almost no extinction have a SED consistent with an on-going star formation over 
some Myrs. We derive a mean dust attenuation law   in the wavelength range 
965-1140 $\rm \AA$ by comparing the SED of 
obscured galaxies to an empirical dust-free SED. The extinction is nearly 
constant longward of 1040 $\rm \AA$ but rises at shorter wavelengths. We compare 
our results with other studies of the extinction for galaxies and stars in this 
wavelength range.\\ 
       }
\titlerunning{FUV SED of Starburst galaxies}
   \maketitle

\section{Introduction}

 The spectral energy distribution (SED) of star-forming galaxies in the far-UV
 (900-2000 $\rm \AA$)  
 is known to be determined by the stellar initial mass function,
 the recent star formation history and dust extinction. 
 As a consequence, the far-UV observations of galaxies are of  
 fundamental importance to know the very recent star formation history in 
 the universe and to interpret the spectra of high-z galaxies. \\

 Thanks to the spectral observations of the {\it IUE} satellite together with 
wide 
 field UV imagers (SCAP (Donas et al. \cite{donas87}) , FOCA 
(Milliard 
et al. \cite{milliard94}, UIT  (Stecher et al. \cite{stecher97})  or FAUST 
 (Deharveng et al. \cite{deharveng94}) experiments), our knowledge of the 
 extinction longward of 1200 $\rm \AA$ has been considerably improved during the 
 last 10 years. 
 Even, if we are far from a complete understanding of the interplay of dust 
 and UV emitting sources within galaxies, 
 empirical laws have been found which allow us to estimate the 
 extinction in galaxies, especially in starbursting objects
 (e.g. Calzetti et al. \cite{calzetti00} and references therein).
 These successes may be traced back to the fact that the UV energy distribution 
 for young starbursts are very similar. They can be fitted by a power-law 
(Leitherer \& Heckman 
 \cite{leitherer95}) and  
 changes in the exponent of the power-law can be attributed to reddening  
(Calzetti et al. \cite{calzetti94}, Meurer et al. \cite{meurer95}).\\
 
 Stellar population models show that    
 these properties  slightly change in the  900 - 1200 $\rm \AA$ domain.  
 The time scale for reaching the equilibrium of UV flux 
 in constant star formation gets shorter and, alternatively, age effects get 
more 
 significant for instantaneous bursts. The  
 SED cannot be fitted by a simple power-law, making the potential 
 separation of age and
 reddening more difficult in practical terms.  
 Observations of the spectral energy distribution of 
 star-forming galaxies downward of 1200 $\rm \AA$  
 have been scarce so far and until recently were limited to those 
 obtained with the Hopkins Ultraviolet Telescope (HUT): 19 spectra have been 
recently analyzed 
 and an attenuation law for star-forming galaxies is derived from 900 to 1800 
$\rm 
\AA$ 
 by Leitherer et al. (\cite{leitherer02}).

  {\it FUSE} ({\it Far Ultraviolet Spectroscopic Explorer})
 has recently opened again an access in the 900-1200 $\rm \AA$ range 
 and the possibility of observing star-forming galaxies in this domain
 (e.g. Thuan et al. \cite{thuan02}, Heckman et al. \cite{heckman01a}). 
 Although its very high spectral resolution is suited to
 the analysis of spectral lines,  {\it FUSE}  can also be 
 used to analyse the  continuum emission of starburst galaxies in an 
 attempt to study their star formation history and their 
 internal extinction. The high spectral resolution becomes an advantage for a 
 better evaluation of the continuum in the presence of numerous absorption 
features as compared to the earlier work of Leitherer et al. 
(\cite{leitherer02}) with HUT data. 
 In the following, we report such an analysis for a sample of 12 starburst 
galaxies.

\section{Observations and data reduction}

\subsection {The sample}

 From June 2000 to December 2001, we  obtained  spectra of the center 
of six nearby starburst galaxies with 
{\it FUSE} (Guest Investigator Program A023). One object was not retained for 
analysis 
 but we added data from  the {\it FUSE} 
archive and 
 built a sample of 12 
starburst galaxies appropriate for studying the FUV (900-1200 $\rm \AA$) 
spectral 
energy distribution.

The starburst galaxies were primarily selected to cover a range as large as
possible in dust extinction.  The dust extinction was estimated with both 
the far-infrared to ultraviolet flux (1600 $\rm \AA$) ratio  and the shape of 
the UV
continuum longward of 1200 $\rm \AA$.  All the targets were detected by {\it 
IRAS} and integrated fluxes
are available at 60 and 100 $\mu$m except for Tol1247-232 which is only detected
at 60 $\mu$m.  The individual fluxes are listed in Table 1.  The FIR (40-120
$\mu$m) emission was calculated by combining the fluxes at 60 and 100 $\mu$m
(Helou et al.  \cite{helou88}).  In the case of Tol1247-232 we adopted the 
average
value  $\rm F_{100}/F_{60}=1.5 $ found for the other galaxies of our sample.  
Getting a FIR to
UV flux ratio representative of the central region observed by {\it FUSE} 
($30\times 
30$ arcsec$^2$) is an issue since the {\it IRAS} data are
integrated over the galaxies. All the galaxies were observed by {\it IUE} but
also through a small diaphragm ($10 \times 20$ arsec$^2$).  In the case of small
galaxies like MRK54, Tol1247-232 or Tol1924-416, the flux at 1600 $\rm \AA$ 
measured
by {\it IUE} can be considered as representative of the total flux emitted by 
the
galaxy and a reliable FIR to UV flux ratio can be estimated.  Unfortunately most
of our galaxies have angular sizes much larger than the {\it FUSE} and {\it IUE} 
apertures.
Therefore we  compiled integrated UV fluxes from the literature which were  
extrapolated to 1600 $\rm \AA$ using the power law $\rm F_{\lambda} \propto
\lambda^\beta$ (Calzetti et al.  \cite{calzetti94}).  For three targets
(NGC3504, NGC7714, NGC1140) no complementary UV data were found.  Nevertheless,
because of the moderate size and the high compactness of NGC7714 and NGC1140 we
 used the {\it IUE} fluxes and the resulting FIR to UV flux ratio is taken 
as an upper limit for these two targets.  No FIR to UV flux ratio was 
calculated for NGC5236:  an integrated FIR to UV flux ratio for this very large
galaxy cannot be representative of the central region observed by {\it FUSE}.

The cases of NGC3690 and MRK59 deserve further comments:\\
NGC3690 is refered as a galaxy pair in the NASA/IPAC Extragalactic Database 
(NED), NGC3690 and IC694  form the Galaxy Triplet Arp299 : the {\it  FUSE} 
aperture is centered on the first 
galaxy of 
the pair referenced as NGC3690 NED1 in 
(NED) whereas {\it IUE}  observed the central position of the 
pair. Meurer et al. (\cite{meurer95}) found a larger flux with {\it HST}/FOC on 
the 
same position as {\it IUE} and concluded that the {\it IUE} aperture was not 
centered on 
NGC3690. Therefore we  adopted the total flux of {\it HST}/FOC and the slope of 
the continuum measured by {\it HST}/FOS (Meurer et al. \cite{meurer95}). 
Nevertheless, 
given the extent of both galaxies of the pair and the fact that  {\it IRAS} 
observed the 
pair as a whole the $\rm F_{FIR}/F_{UV}$ ratio is probably overestimated and has 
been 
flagged as an upper limit. Moreover since 
the {\it FUSE} aperture is centered on one galaxy of the pair, one must be 
cautious in  
comparing the observations of {\it FUSE}, {\it IRAS} and {\it HST}. \\
MRK59 is an HII region observed as a bright knot in the southern part of 
NGC4861. 
Thuan et al. (\cite{thuan02})  used the {\it FUSE} spectrum to study the heavy 
element abundance and the H$_2$ content of MRK59.
Both {\it FUSE} and {\it IUE} apertures were centered on MRK59. The {\it IRAS} 
detection is also at the position of MRK59 but given its low resolution the FIR 
flux concerns NGC4861. NGC4861 has been observed in UV by the SCAP 
experiment at 2000 $\rm \AA$ and its total flux is a factor of $\sim 2$ larger 
than that 
measured within the {\it IUE} aperture. The FIR to UV flux ratio is calculated 
with 
the total UV flux.\\

\begin{table*}
\caption[]{The sample of starburst galaxies with some useful characteristics. 
FIR fluxes at 60 and 100 $\mu$m (F(60) and F(100)), galaxy sizes, Galactic 
extinctions and redshifts come from NED.
The 
total observed UV 
flux at 1600 $\rm \AA$ (column 7) and the slope $\beta$ of the continuum ($\rm 
F_{\lambda} = \lambda^{\beta}$) (column 4) come from various sources. a: Meurer 
et al. 
\cite{meurer99}; 
b: NED; c: Bell \& Kennicutt \cite{bell01};
d: Meurer et al. \cite{meurer95} (the UV flux at 1600 $\rm \AA$ is extrapolated 
from 
2200 $\rm \AA$ with the law $\rm F_{\lambda} = \lambda^{\beta}$); e: Donas et 
al. 
\cite{donas87} (the UV flux at 1600 $\rm \AA$ is extrapolated from 
2000 $\rm \AA$ with the law $\rm F_{\lambda} = \lambda^{\beta}$); f: Kinney et 
al.  
\cite{kinney93}; g: Terlevich et al. \cite{terlevich93}; h:  Giavalisco et al. 
\cite{giavalisco96}. The observed UV 
fluxes  were corrected for the 
Galactic extinction before the calculation of $\rm 
F_{FIR}/F(1600)$ (column 8). When the galaxy is very extended as compared to the 
{\it FUSE} 
aperture,  $\rm 
F_{FIR}/F(1600)$ is  flagged with an asterix. The metallicities  (column 
9) are taken 
from Heckman et al. (\cite{heckman98}) except for few cases: Calzetti et al. 
(\cite{calzetti94}) for NGC7673, Noeske et al. (\cite{noeske00}) for MRK59, 
Terlevich et al. (\cite{terlevich93}) for Tol1247-232}.
    
\begin{flushleft}
\begin{tabular}{llcccccccc} 

\hline 
name  & Size &E(B-V)$\rm _{G}$ &$\beta$& F(60) &  
F(100) & F(1600) & $\rm 
\log$& Metallicity& redshift \\ 
      & arcmin$^2$ & mag   & &   Jy &   Jy  & $\rm erg~cm^{-2}~s^{-1}~\AA^{-1}$& 
$\rm (F_{FIR}/F(1600))$ &12+log 
(O/H)&z \\
       
\hline 
NGC1140 &$1.7\times 0.9$&0.038&-1.78$^a$&3.36  &4.92  &$ 4.2~10^{-14}$$^a$  
&$<$ 
0.28&8.0&0.005\\
NGC1705 &$1.9\times 1.4$&0.008&-2.42$^a$&0.87  &1.81  &$25.4~10^{-14}$$^b$  
&-0.93&8.0&0.002\\
NGC3310 &$3.1\times 2.4$&0.022&-1.05$^a$&34.13 &47.95 &$52.2~10^{-14}$$^c$  &  
0.24*&9.0&0.003\\
NGC3504 &$2.7\times 2.1$&0.027&-0.56$^a$&22.70 &35.70 &                   &     
& 9.1&0.005\\
NGC3690 &$1.2\times 1.0$&0.017&-0.74$^d$&121.64&122.45&$3.73~10^{-14}$$^d$& 
$<$1.9& 8.8&0.010\\
NGC5236 &$12.9\times 11.9$&0.066&-0.83$^a$& 266& 639  &$ 233.3~10^{-14}$$^c$&     
 & 9.3&0.002\\
NGC7673 &$1.3\times 1.2$&0.043&-1.50$^a$&4.91  &6.89  &$7.5~10^{-14}$$^e$& 
0.19& 8.5&0.011\\
NGC7714 &$1.9\times 1.4$&0.052&-1.23$^a$& 10.52&11.66 &$3.3~10^{-14}$$^a$& 
$<$0.81& $>8.7$&0.009\\
MRK54   &$0.7\times 0.4$&0.015&-1.78$^a$& 0.93 &1.83  &$1.7~10^{-14}$$^a$& 
0.25& &0.045 \\
MRK59   & HII region    &0.011&-2.57$^f$& 1.97 &2.46  
&$19.9~10^{-14}$$^e$&-0.56*& 8.0&0.003\\
Tol1247-232&            &0.089&-1.25$^h$& 0.50  &      
&$0.9~10^{-14}$$^g$&-0.03& 8.1&0.048\\
Tol1924-416&$0.8\times 0.4$&0.087&-2.12$^a$&1.68&1.01 
&$3.3~10^{-14}$$^a$&-0.18& 8.1&0.010\\
\hline 
\end{tabular} 
\end{flushleft} 
\end{table*}

\subsection{Data processing}

All the fluxes were recorded with the LiF and SiC channels 
through the LWRS ($30\times 
30$ arcsec$^2$) aperture.  A summary of the observations is given in Table 2. 
Each 
observation consisted of several exposures. 
 For the purpose of inter-comparison we  report in Table 2 the signal to 
noise ratio 
for one spectral window 
used to 
measure the SED (window 4 at 1050 $\rm \AA$). The S/N values are rather low 
because 
of the 
small size of the window (9 $\rm \AA$).\\
The data were prepared for analysis by the {\it FUSE} data pipeline. Since the 
observations spanned over 18 months several versions of the pipeline were  used 
to reduce the data. We  compared the different versions  of the pipeline to the 
last version (2.0.5) available at the time of writing this paper. The wavelength 
calibration has slightly changed; nevertheless since  we are only concerned by 
the  continuum, this change does not affect our results. We also observed a 
 change in the absolute fluxes  especially between the earliest versions  
 of the pipeline and the last one but once  again this modification does not 
affect  the shape of the continuum. Since we are only interested by relative 
fluxes  we have not entirely re-processed the data with the version 2.0.5 of the 
pipeline. For the purpose of information however, we list 
in 
Table 2 the approximative multiplicative factor  to obtain  absolute 
fluxes consistent with the last version of the pipeline.\\

We  combined the individual exposures for each target weighted by the respective 
exposure 
times after checking that the mean fluxes and errors were  consistent in the 
spectral windows of interest (see below for the definition of the windows). In 
few cases some exposures were discarded because of discrepancies between the  
mean fluxes not consistent with the errors. \\
For each spectral window we then combined the 
two segments with the highest sensitivity weighted by their effective area taken 
from the {\it FUSE} observer's guide. Once again the combination was made after 
checking that the mean 
fluxes and errors were consistent in each individual segment: the inconsistent 
data 
 were discarded. In particular, the fluxes collected on the LiF2 channel  were 
found 
lower than those obtained on the LiF1 channel for three targets (Tol1247-232, 
Tol1924-416 and MRK59). This effect is probably due to  the target 
moving outside the LiF2 spectrograph aperture;   in these cases we have only 
considered the LiF1 channel. 
We  checked that the SiC channels were not affected by this effect. \\

We  compared the fluxes measured by {\it FUSE} at 1180 $\rm \AA$ with the 
spectra obtained by {\it IUE} longward of 1200 $\rm \AA$. The fluxes were 
consistent 
within 
30$\%$, which seemed reasonable given the different apertures  (the fluxes 
are compared without any correction factor for the apertures) and the low 
sensitivity of {\it IUE} near 1200 $\rm \AA$. The only discrepant case was 
NGC3690 
for 
which {\it FUSE}  observed a flux more than twice that measured by {\it IUE}. As 
discussed in the previous section, the discrepancy is likely to be due to the 
bad centering of {\it IUE}.

\begin{table}
\caption[]{Summary of the observations. The factor listed in the last column is 
the mean ratio between the last absolute flux calibration (pipeline version 
2.0.5) and the flux calibration effectively used to reduce the data, see text 
for details}

\end{table}

\begin{flushleft}
\begin{tabular}{lllll} 

\hline 
Target&$\rm T_{exp}$ &S/N  &pipeline &factor\\
 name  &sec & 1050 $\rm \AA$ &version& \\
 NGC1140 & 4033  & 11  &1.8.7&0.94\\
 NGC1705& 7658 & 11 &1.6.6&1.20 \\
 NGC3310& 27077 & 10&1.6.9&1.20\\
 NGC3504 & 13561& 7& 1.8.7&0.94\\
 NGC3690 & 15957 & 8.5 &1.8.7&0.94\\
 NGC5236& 26538 & 7.5&1.7.7&0.97\\
 NGC7673& 10515 & 5.5 &1.7.7&0.97\\
 NGC7714 & 5796 & 5.5&1.7.7&0.97\\
 MRK54  & 27502 & 4&1.6.8&1.20\\
 MRK59  & 9831 & 9 &1.6.6&1.20\\
 Tol1247-232& 24303 & 5&1.6.9&1.20\\
 Tol1924-416& 5014 & 5.5 &1.6.9&1.20\\
\hline 
\end{tabular} 
\end{flushleft}

\section{ Deriving the  Spectral Energy Distributions} 

Our aim is to extract the spectral energy distribution  from the very high 
resolution {\it FUSE} spectra. We have to get rid of the numerous absorption 
 and emission lines from the Milky Way and the airglow. Given the low 
sensitivity at the shortest wavelengths and  the presence of  
  the Lyman series of  HI we do not consider wavelengths shorter than 960 
$\rm \AA$.\\

The adopted strategy is to define spectral windows avoiding  Galactic   
 and airglow lines: we are able to define 13 windows.  They are 
described in Table 3. For the interstellar  absorption lines 
we  use the list of Barnsted et al. (\cite{barnstedt00}). Two windows (3 and 
4) contain   weak  Galactic
interstellar lines (SiII 1020.7$\rm \AA$  and  FeII 1055.5$\rm \AA$), for each 
target we  
check that these lines are not present in the spectra. For the first window 
we 
 use either  window 1 or 1b according to the redshift of the target and 
the position of the lines. We have also to deal 
with  potential absorption lines of Galactic molecular hydrogen  along the line 
of sight. The windows that we define are taken as much 
as 
possible outside the Lyman and Werner bands of H$_2$. Nevertheless some lines 
remain. The contamination induced by this effect is estimated both with 
the {\it FUSE} 
simulator and with observed reference spectra (Shull et al. \cite{shull00}) and 
does not 
exceed a few 
percents. \\ 
\begin{table}

\caption[]{Spectral windows for the determination of the SED. Window 1 contains  
three segments 969-971, 974-976 and 979-981 $\rm \AA$. For 
each 
target we  used either  window 1 or  window 1b  according to the 
position of the absorption lines}
    
\begin{flushleft}
\begin{tabular}{llll} 

\hline 
Num.   & $\Delta \lambda$ &$\lambda_{eff}$&detector\\
      & $\rm \AA$    &  $\rm \AA$ & segment  \\
  1 & 969-971 974-976 979-981&975 &SiC2A SiC1B \\
 1b & 967-971&969 &SiC2A SiC1B\\
  2 & 996-1001 & 999&SiC2A LiF1A \\
  3 & 1016-1024 & 1020&LiF1A LiF2B \\
  4 & 1052-1061 & 1057 &LiF1A LiF2B \\
  5 & 1068-1076&  1072&LiF1A LiF2B\\
  6 & 1087-1091 &1089& LiF2A \\
  7 & 1100-1107 &1104&  LiF2A LiF1B\\
  8 & 1113-1120&1117& LiF2A LiF1B \\
  9 & 1126-1130& 1128&LiF2A LiF1B \\
 10 & 1136-1143& 1140&LiF2A LiF1B \\
 11 & 1146-1152& 1149&LiF2A LiF1B \\
 12 & 1155-1167& 1161&LiF2A LiF1B \\
 13 & 1170-1175 &1173& LiF2A LiF1B \\    
\hline 
\end{tabular} 
\end{flushleft} 
\end{table}
It is a much more difficult task to avoid the absorption lines in the objects 
themselves  
because their position varies with the redshift of the observed galaxy. Each 
target is analysed separately. If a  window contains  a line from the Lyman 
series of HI it is suppressed. For the metal lines, if they are located at the 
edge of the window,  we  
 avoid them  by redefining the limits of the window. For the lines located 
well within a window two cases are found: if the line is detected (or expected 
to 
be 
strong when the spectrum has a low S/N)   the window is suppressed or if the 
line is undetected (or expected to be weak when the spectrum has a low S/N)  
the window is  kept.\\

The molecular hydrogen in the target may also cause some problems. 
Nevertheless, no H$_2$ is  detected in the spectra of our galaxies, specifically 
the Lyman 4-0 and 3-0 bands are not 
 detected even in the best spectra. Such an 
absence of H$_2$ lines was also reported for I Zw18 (Vidal-madjar et al. 
\cite{vidal00}) and two galaxies of our sample NGC1705 (Heckman et al. 
\cite{heckman01a}) and MRK59 (Thuan et al. \cite{thuan02}). We  
assume
that no H$_2$ line 
contaminates the continuum of the galaxies even when the H$_2$ lines may not 
be detected because of a low S/N ratio.\\ 

The fluxes obtained in each spectral window are reported in Table 4. As 
discussed in section 2.2 different versions of the {\it FUSE} pipeline were used 
to obtain these absolute fluxes. The mean  factor to apply in order to obtain   
an absolute  calibration consistent with the  latest version of the pipeline 
(version 2.0.5) is given in Table 2. 
The errors quoted are estimated as the dispersion of the measurement within a 
spectral window (i.e a combination of exposures and detector segments). It  is  
checked that the error is always larger than the Poissonian error calculated 
by the 
{\it FUSE} pipeline.

   \begin{figure*}
  \includegraphics[angle=-90,width=18cm]{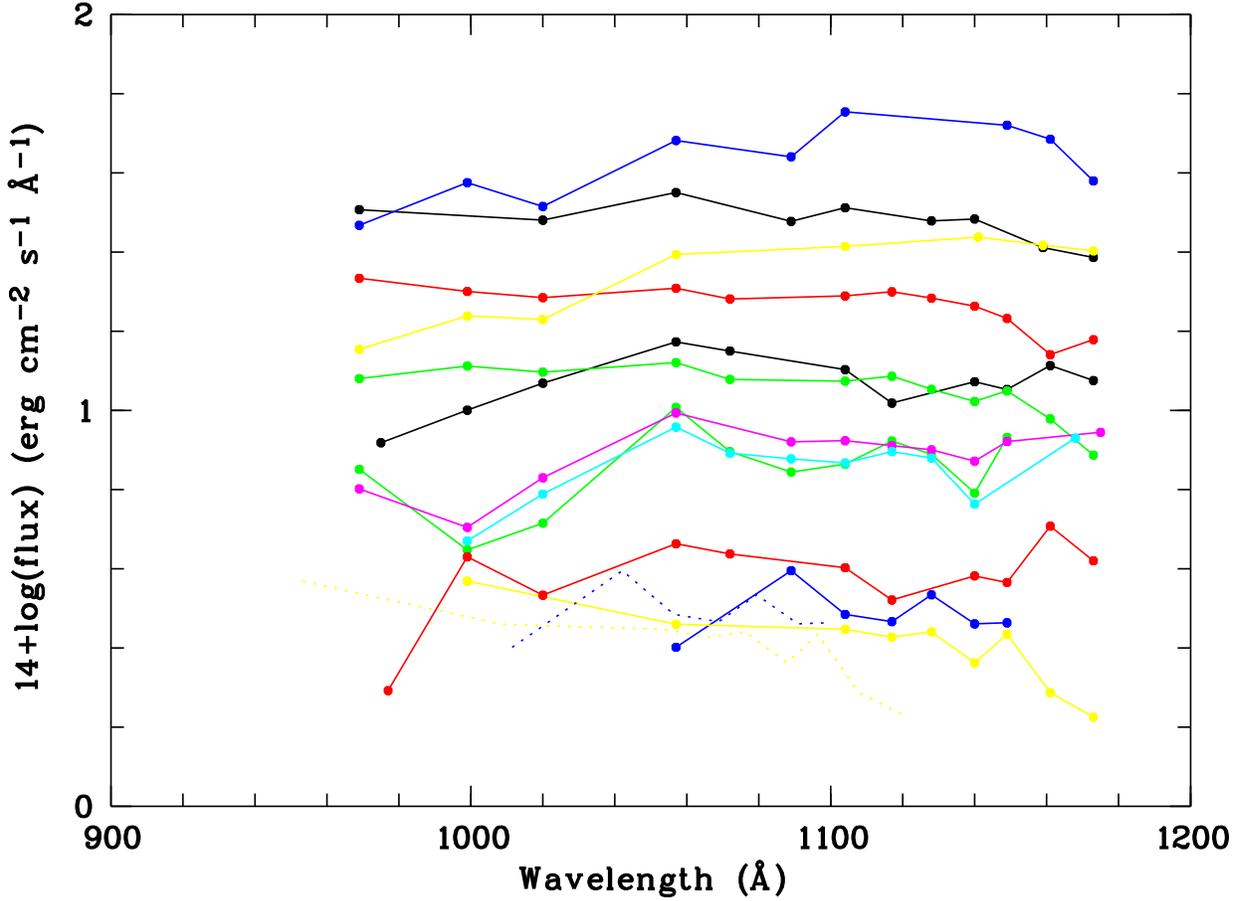}
   \caption{FUV Spectral Energy Distributions. The fluxes were corrected 
for the Galactic extinction. By increasing flux at the largest observed 
wavelength: 
Tol1247-232 (yellow), MRK54 (blue), NGC3504 (red), Tol1924-416 (green), NGC7673 
(cyan), NGC7714 (magenta), NGC3690 (green), NGC1140 (black),    
 MRK59 (red), NGC1705 (black), NGC3310  (yellow) and NGC5236 (blue). 
   For the two distant galaxies Tol1247-232 and MRK54, the rest-frame 
wavelength corrected distributions are plotted with dotted lines. }
      \label{sedcompil}
         \end{figure*}
         
The SED of the galaxies, as defined by the fluxes in the spectral windows, are 
plotted in  Fig.~\ref{sedcompil}.  The 
fluxes 
 are  corrected for the dust extinction inside our Galaxy.  The color excess 
along the line of sight are from Schlegel et al. (\cite{schlegel98}). We  adopt 
the extinction curve deduced by Sasseen et al. (\cite{sasseen02}). The color 
excess is low along the line of sight of the targets (cf. Table 1) and the 
correction is not large. We  checked that the resulting SED are not 
sensitive to 
this correction by repeating the calculation  using an extrapolation of   the 
extinction curve of 
Cardelli et al. 
(\cite{cardelli89}) down to 900 $\rm \AA$. 
We  
report the observed wavelengths except for the two most distant galaxies 
(MRK54, Tol1247-232) for which rest frame wavelengths are considered.

\begin{table*}
\caption[]{The observed Spectral Energy Distribution of the sample galaxies. The 
  fluxes are in $\rm 10^{-14} erg~cm^{-2} s^{-1} \AA^{-1}$.
  For each galaxy, the first line corresponds to the observed fluxes estimated 
in the spectral windows defined in Table 3, the second line is the error on the 
estimation of the flux. No Galactic extinction correction was applied.
  When the galaxy name has an asterix $*$ the first column corresponds to the 
wavelength 969 $\rm \AA$ (spectral window 1b) instead of 975 $\rm \AA$ (spectral 
window 
1).The ratio R equal to F(1070 $\rm \AA$)/F(1010 $\rm \AA$) is listed in the 
last 
column together with its estimated error in the second line. Note that the 
fluxes were corrected for Galactic extinction before the 
calculation of R.}

\begin{flushleft}
\begin{tabular}{lllllllllllllll} 

\hline 

Galaxy   &975$^*$ &999 &1020 &1057 &1072 &1089 &1104 &1117 &1128 &1140 &1149 
&1161 
&1173& R\\ 
\hline 
 &&&&&&&&&&&&&\\
NGC1140     &4.9 &6.1 &7.3 &9.6 & 9.2 &     &8.4 &7.0   &    &8.0 &7.7 &8.9 
&8.2&1.3\\
            &1.7 &1.7 &1.7 &0.9 & 0.9 &     &0.9 &0.9   &    &0.9 &0.9 &0.9 
&0.9&0.3\\
NGC1705$^*$ &28.8&     &27.4&32.4&      &27.5&29.8&     &27.7&28.1&     
&23.8&22.5&1.1\\
            &3.0 &     &3.0 &3.0 &      &3.0 &3.0 &     &3.0 &3.0 &     &3.0 
&3.0&0.2\\                    
NGC3310     &10.5&13.0 &12.9&19.2&      &    &20.5&     &    &21.9&     
&21.0&20.4&1.5\\
            &3.0 &3.0  &3.0 & &      &    & &     &    & &     & 
&&0.3\\
NGC3504     &1.4 &3.0  &2.4 &3.4 &3.2   &    &3.0 &2.5  &    &2.9 &2.8  &3.9 
&3.2&1.2\\
            &0.7 &0.7  &0.5 &0. 5&0.4   &    &0.4 &0.4  &    &0.4 &0.4  &0.4 
&0.4 &0.3  \\          
NGC3690$^*$ &5.6 &3.6  &4.2 &8.4 &6.5   &5.8 &6.1 &7.0  &6.5 &5.2 &7.2  &9.2 
&8.4&1.7\\
            &1.3 &1.3  &1.3 &1.0 &1.0   &1.0 &1.0 &1.0  &1.0 &1.0 &1.0  &1.0 
&1.0&0.6\\
NGC5236$^*$ &11.7&15.9 &14.4&22.4&      &21.2&28.0&     &    &    &27.0 
&25.1&19.9&1.3\\       
            &3.0 &3.0  &3.0 &3.0 &      & & &     &    &    &  & 
&&0.3\\
NGC7673     &    &2.7 &3.6 &5.5 &4.8   &4.7 &4.7 &5.0  &4.9 &3.7 &     &    
&5.6&1.5\\            
            &    &1.0  &1.0 &1.0 &1.0   &1.0 &1.0 &1.0  &1.0 &1.0 &     &    
&1.0&0.6\\
NGC7714$^*$ &3.1 &2.6  &3.5 &5.4 &      &4.7 &4.8 &4.7  &4.7 &4.4 &4.9  &    
&5.3&1.5\\
            &1.1 &1.1  &0.8 &0.8 &      &0.8 &0.8 &0.9  &0.8 &0.8 &0.9  &    
&0.8&0.6\\
MRK54       &    &     &    &2.1 &      &3.3 &2.6 &2.5  & 2.9&2.5 &2.5  &    &   
&1.3\\        
            &    &     &    &0.5 &      &0.3 &0.3 &0.3  &0.3 &0.3 &0.3  &    &   
&0.3\\
MRK59$^*$   &18.5&17.3 &16.8&17.9&16.9  &    &17.3&17.8 &17.2&16.4&15.3 
&12.4&13.6&1.0\\
            & &  & & &   &    & &  & & &  & 
&&0.2\\         
Tol1247-232    &    &1.2  &    &1.0 &      &    &1.1 &1.1  &1.1 &0.9 &1.1  &0.8 
&0.7&0.9\\
               &    &0.5  &    &0.2 &      &    &0.3&0.3   &0.3 &0.3 &0.3  &0.3 
&0.3&0.2\\
Tol1924-416$^*$&3.6 &4.2  &4.2 &4.8 &4.5   &    &4.7&4.9   &4.6 &4.3 &4.7  &4.0 
&3.3&1.0\\
               &1.3 &1.0  &1.0 &0.9 &0.9   &    &0.9&0.9   &0.9 &0.9 &0.9  &0.9 
&0.9&0.3\\           
\hline 
\end{tabular} 
\end{flushleft} 
\end{table*}

\section{Analysis of the far-UV Spectral Energy Distributions}

\begin{figure}
  \includegraphics[width=8cm]{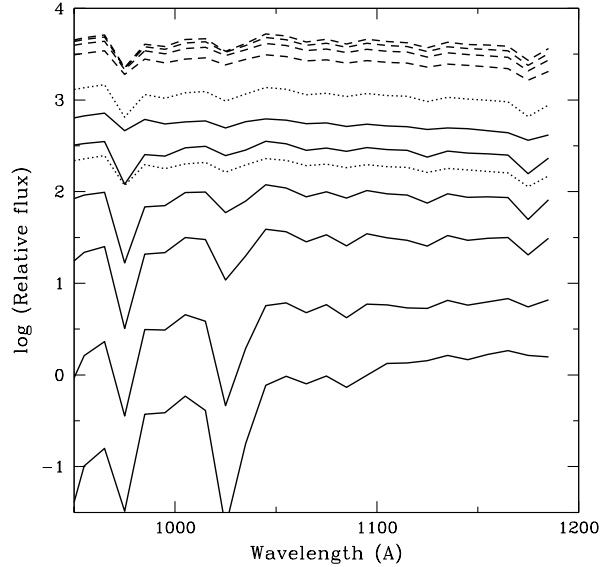}
   \caption{The SED predicted in the far-UV by the population synthesis model 
     Starburst99. Solid lines show the flux decrease of an instantaneous burst 
      (1 M$_{\sun}$)
     observed after 1, 5, 10, 20, 50 and 100 Myrs. Dashed lines show 
     a constant star formation rate (10$^6$ M$_{\sun}$ yr$^{-1}$) 
     lasting (from bottom to top) 5, 10, 20 
     and 50 Myrs. A Salpeter IMF (slope $-$2.35) is adopted from 1 to 100 
     M$_{\sun}$  and the metallicity is solar. Dotted lines  show the 
impact of a truncation of the IMF at 30 
     M$_{\sun}$ for the instantaneous  burst of 1 Myr (bottom) and for the 5 
Myrs
     continuous star formation (top). While the latter difference persists as a 
function 
     of time (not shown for clarity), the impact becomes hardly 
     noticeable for bursts older than a few Myrs; in both cases the relative 
     spectral shape is not changed very much.}
      \label{modelsb99}
      \end{figure}
      
 \begin{figure}
  \includegraphics[angle=-90,width=8cm]{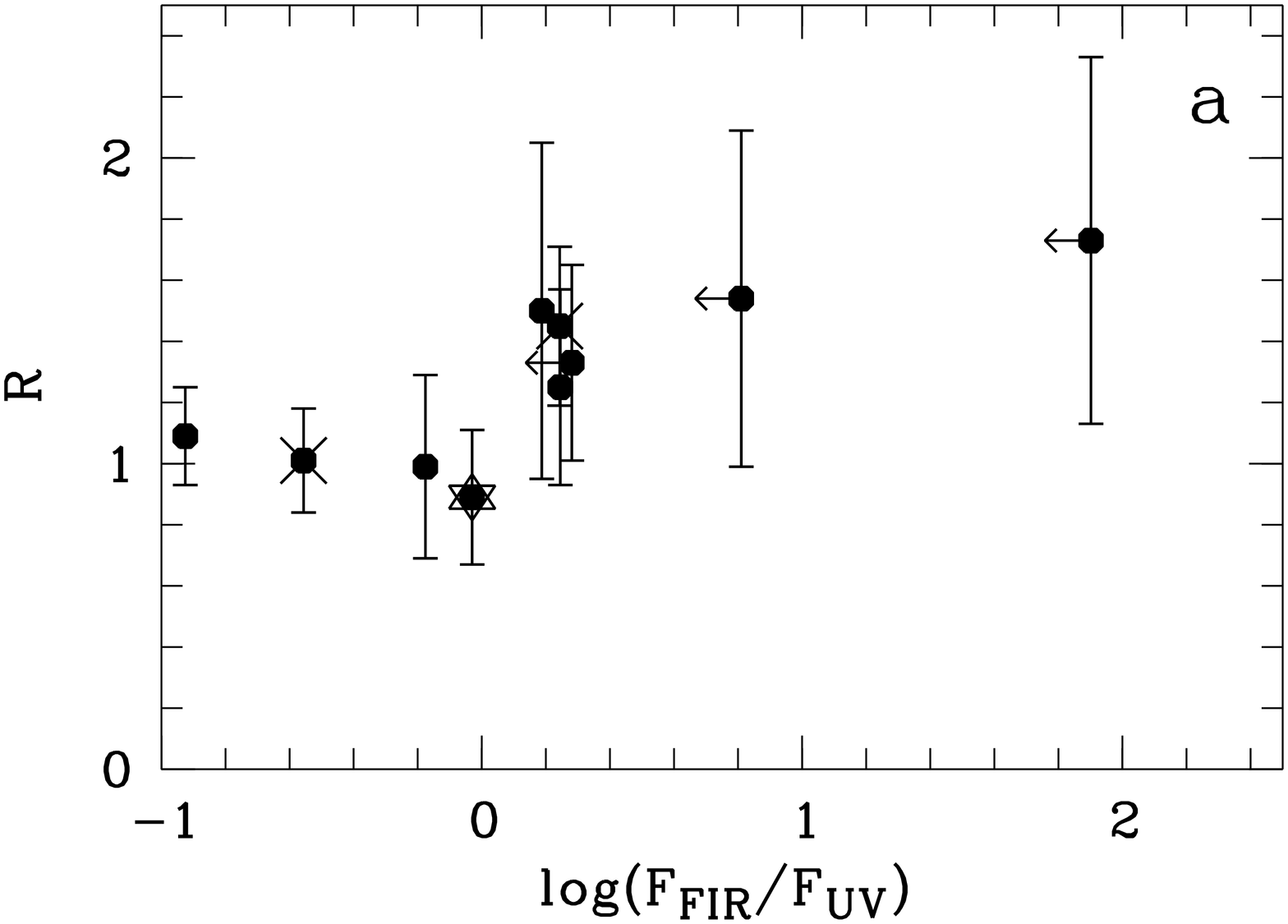}
  \includegraphics[angle=-90,width=8cm]{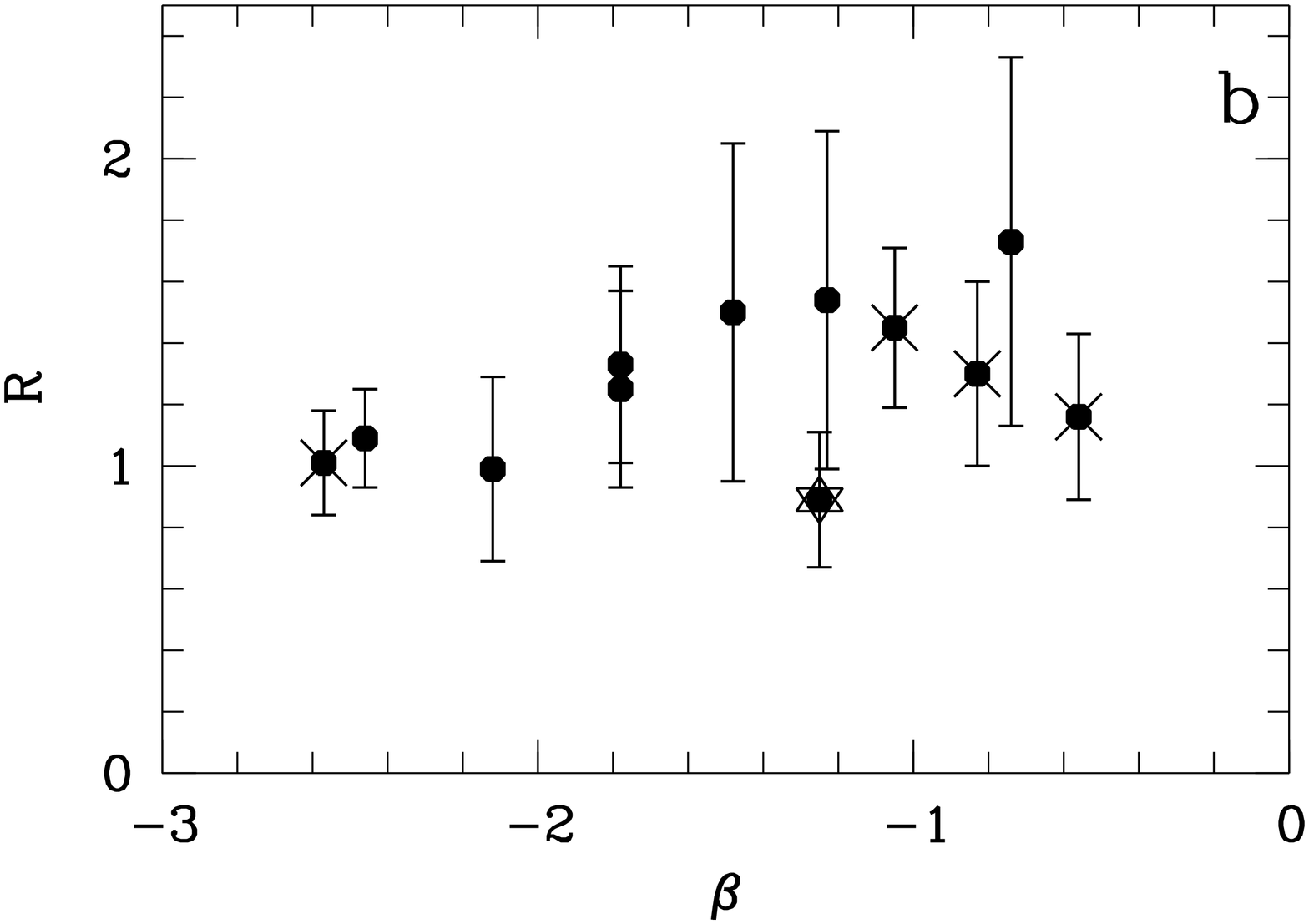}
     \caption{R = F(1070 $\rm \AA$)/F(1010 $\rm \AA$) plotted against $\rm 
\log(F_{FIR}/F_{UV})$ ({\bf a}) and the slope $\beta$ of the UV continuum (
{\bf b}). Extended 
galaxies 
 for which {\it FUSE} and {\it IUE} data only concern the central part are 
plotted with 
crosses (NGC3310, MrK59, NGC3504 and NGC5236). Tol1247-232 is 
plotted with a star.}
      \label{ratiofirfuvbeta}
         \end{figure}     
  The first step to understand the basic features of the SED displayed in 
  Fig.~\ref{sedcompil}, a rather flat 
  distribution from 1180 \AA~ down to $\sim$ 1050 \AA~ and a more or less 
  pronounced decrease at wavelengths shortward of 1050 \AA,
  is to compare them with the predictions of 
  population synthesis.
  The Starburst99 models (Leitherer et al. \cite{leitherer99}) are especially
  appropriate for this comparison and are presented in 
  Fig.~\ref{modelsb99} at different steps in time. 
  Fig.~\ref{modelsb99} shows the rapid decline of the flux (especially at short 
  wavelengths) for an instantaneous burst and, correlatively, the rapid 
equilibrium 
  reached in continuous star formation regime.  

    The four galaxies with a relatively flat SED (or slightly increasing at 
short
  wavelengths) in  Fig.~\ref{sedcompil}, NGC1705, MRK59, Tol1924-416 and 
  Tol1247-232  can be
  easily interpreted as a recent (age $\leq$ 10 Myrs) instantaneous starburst
  or a continuous star formation. 
  The spectral shapes are the same because of 
  the short time scale
  between star formation and star evolution off the domain of high effective
  temperatures. The lack (or small amount) of dust extinction is consistent with 
the 
  low values of $\beta$ for four of these objects (MRK54 is intermediary while 
the 
case
  of Tol1247-232, $\beta = - 1.2$, will be rediscussed later on). 

    The interpretation of the eight other objects is not as straightforward  
because the 
  decrease of the flux at short wavelengths may result from one or a combination 
of 
  two factors: an aging burst 
  (or more complex star formation history) or dust extinction (in general 
  reddening is known to increase at short wavelengths). 
  The metallicity is known to be only a second-order effect 
  and variations of the IMF within reasonable limits do not alter the 
  spectral shapes. Although a separation of age and reddening 
  is difficult, we present several arguments in the following
  for a dominant role of dust extinction.

\begin{figure}
  \includegraphics[angle=-90,width=8cm]{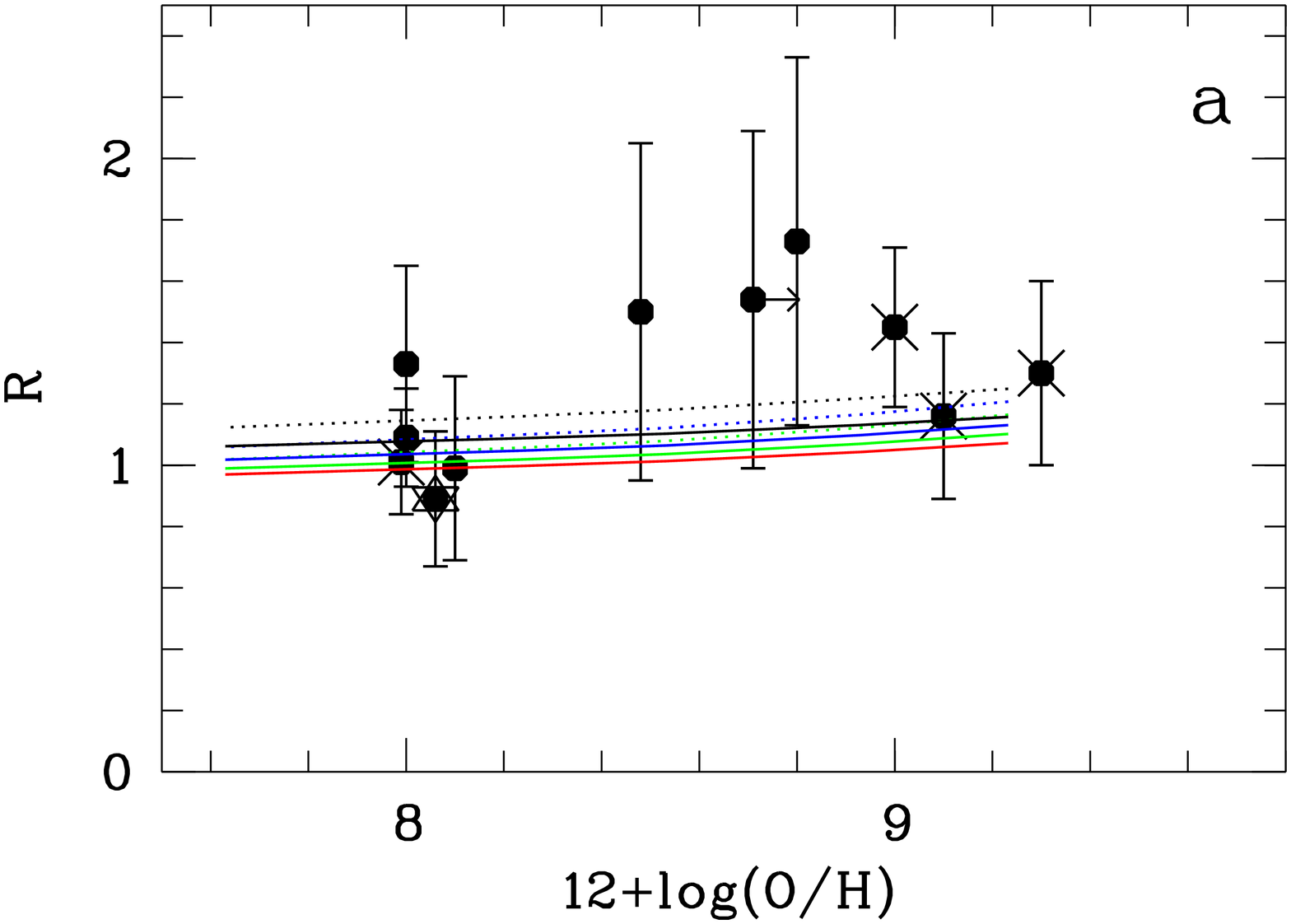}
  \includegraphics[angle=-90,width=8cm]{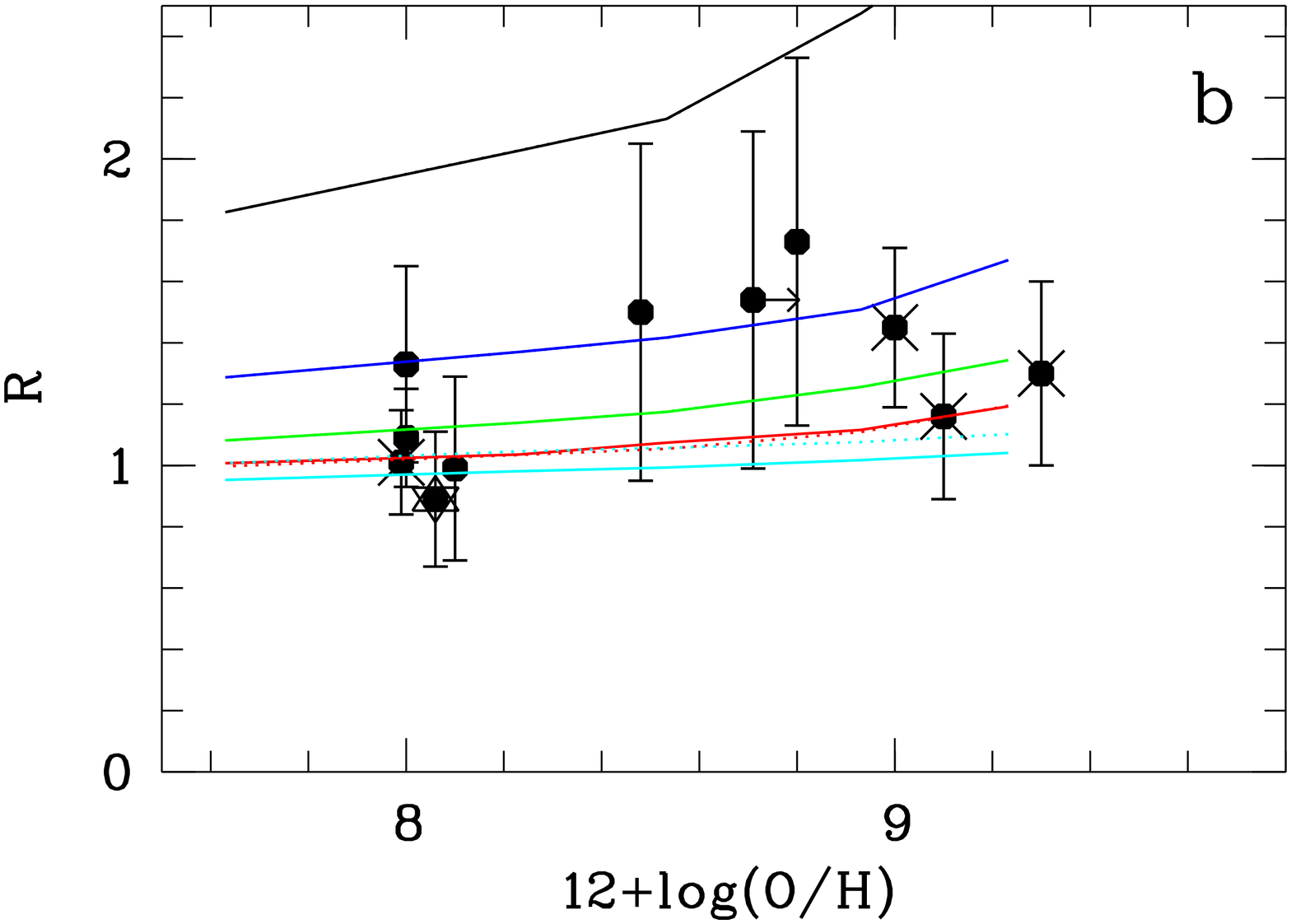}
   \caption{R = F(1070 $\rm \AA$)/F(1010 $\rm \AA$) plotted against the 
metallicity of 
the galaxies. The symbols are the same as in Fig.~\ref{ratiofirfuvbeta}. The  
predictions of the models for ({\bf a}) a constant star formation  over 5 (red), 
10 (green), 
20 (blue) and 50 (black)  
Myrs and ({\bf b}) 1 (cyan), 5 (red), 10 (green), 20 (blue) and 50 (black) Myrs 
after an 
instantaneous burst  are 
overimposed. A Salpeter IMF has been adopted from 1 to 100 
M$_{\odot}$) (solid line) or truncated at 30 M$_{\odot}$ (dashed line)  }
      \label{ratio_metallicity}
         \end{figure}

\begin{enumerate}

  \item The entrance aperture of {\it FUSE} is large enough to smooth the local
variations of the star formation activity, which is likely to result in a
continuous star formation.  In such conditions, only dust extinction can produce
the decrease of the flux observed at short wavelengths.

\item Some extinction is known to be present.  Following Calzetti, Meurer and
   their collaborators the extinction of starburst galaxies at $\rm \sim 1600
   \AA$ can be estimated from their $\rm F_{FIR}/F_{UV}$ ratio or from the slope
   $\beta$ of the UV continuum between 1250 and 2500 $\rm \AA$ and defined as
   $\rm F_{lambda} \propto \lambda^\beta$ (e.g.  Meurer et al.  \cite{meurer99},
   Calzetti et al.  \cite{calzetti00}).  Using the empirical law of Calzetti et
   al.  (\cite{calzetti00}) $$\rm A(1600 \AA) = 2.31 \beta +4.85$$ leads to an
   extinction in the 12 objects that ranges from 0 to 2 mag.

  \item We   develop the previous argument and search whether  
 the decline of the SED at short wavelengths is correlated 
 with some tracers of the dust extinction. To this aim we  choose to quantify 
the
decrease of the SED by averaging the fluxes measured shortward of 1050 $\rm \AA$ 
in  windows  2 and 3 ($\rm \lambda~1010~\AA$) and longward of 1050 $\rm \AA$ 
in 
 windows 4 to 6 ($\rm \lambda~1070~\AA$).  The  ratio of 
the averaged flux at long wavelengths over the averaged flux at short 
wavelengths  
will be named R hereafter. The galaxies with a value of R close to unity have a 
flat spectrum. \\
We compare the values of R with  $\rm 
F_{FIR}/F_{UV}$,  
$\beta$ (Fig.\ref{ratiofirfuvbeta}) and the metallicity of the galaxies 
(Fig.~\ref{ratio_metallicity}). 
In spite of the large error bars, there is an increase of R with $\rm 
F_{FIR}/F_{UV}$; the trends are less pronounced but still present with  
12+log(O/H) and  $\beta$. 
Both $\beta$ and $\rm F_{FIR}/F_{UV}$ are strongly 
correlated with 12+log(O/H) (Heckman et al. \cite{heckman98}), all quantities 
being related to the amount of dust. Therefore the presence of dust 
extinction seems to be related to the decrease of the  far UV SED downward 
$\rm ~1050  \AA$.

 \item In the absence of dust extinction, the interpretation of the decrease 
      of the flux at short wavelengths would require an old starburst.In order 
to  illustrate the comparison of the models 
and the data, we report in 
Fig.~\ref{ratio_metallicity}  the predictions of the Starburst99 models for R.

 A constraint on the age of the stellar population may come from the 
analysis of 
spectral features present in the UV spectra. The UV absorption lines of NGC1705, 
Mrk59, NGC3310 and NGC5236 as observed by FUSE have already been discussed 
(Heckman 
et al. \cite{heckman01a} \cite{heckman01b}, Thuan et al. \cite{thuan02}). We 
focus on the O VI $\lambda$1032 line since it is specifically studied by 
Gonzalez Delgado et al. (\cite{gonzalez97}) as a star formation history 
indicator; the OVI, Ly$\beta$ and CII lines are also modeled with a high 
spectral resolution in Starburst99. Since the analysis requires good S/N ratio
 spectra, we can only add  NGC1140 and NGC3690 to the previously published cases 
listed above. Another limitation is the possible by  contribution of the 
interstellar medium 
for the formation of absorption lines, especially in starburst galaxies 
(e.g. Gonzalez Delgado et al. \cite{gonzalez97}).

OVI $\lambda$1032 is present in the spectra of all the galaxies except for 
NGC3690 for which, unfortunately, an emission line of the airglow coincides with 
the OVI line.   If the origin of this line is stellar,  massive O stars must be 
present and this argues for the presence of a very young star population. In 
case of NGC3690 the shape of the Ly$\beta$ and CII lines is  quite consistent 
with a constant star formation rate over some Myrs or an instantaneous starburst 
not older than 20 Myr. The case of NGC3310 is  puzzling since Leitherer et 
al. \cite{leitherer02} conclude to the lqck of  massive ionizing O stars from 
the HUT spectrum 
of the galaxy. One plausible explanation is that the FUSE and HUT apertures does 
not encompass the same regions of this large galaxy.

Therefore, the analysis of the OVI line supports  a very recent 
star formation in the galaxies and  makes very unlikely  an old 
starburst to explain the decrease of the spectra.

   \item  We  explore for all 12 galaxies how well the fluxes observed within 
each 
      spectral window fit the models in terms of $\chi^2$ minimization.
      We basically use the models displayed in Fig.~\ref{modelsb99} with 
      a full set of metallicity, 12+log(O/H)=8.93 (solar,  Heckman et al. 
\cite{heckman98}), 
      9.23, 8.53, 8.23 and 7.62 in order to reproduce the range of 
      metallicity of the targets (Table 1). 
      We  interpolate 
      the results of the models given by Starburst99 to predict the 
      fluxes within each spectral window defined in Table 3.\\ 
 The 3 galaxies with a value of R close to unity (Tol1924-416, MRK59 and 
NGC1705) 
are very well fitted with a constant star formation rate  over 5 to 20 Myrs 
or  very recent 
instantaneous bursts (1-10 Myrs) and a Salpeter IMF from 1 to 100 
M$_{\odot}$. Indeed, to obtain a low value of R, the 
presence of very massive stars is necessary and an IMF deficient in massive 
stars (with $\rm M_{up} = 30 M_{\odot}$)  leads to too 
large values of R.\\
 In contrast, for the galaxies exhibiting a larger value 
of R (i.e. a decrease of their spectrum at low wavelengths) 
the fits of the 
overall spectra with synthesis models are not good (reduced 
$\chi^2$ always larger than 2) and reproduce only marginally  the 
observed values of R. This poor fitting when only the stellar content is 
accounted for also argues for the role of the dust extinction in re-shaping the 
SED of the galaxies with a large R.
\end{enumerate}

   Before concluding this series of arguments, we come back 
 to the intriguing case 
of Tol1247-232. This galaxy  exhibits the lowest value of R and 
its flux downward of 1200 $\rm \AA$ is 
continuously rising, the best fits are for a very recent 
burst (1 Myr) or a constant star formation over at most 5 Myrs. However, no 
model 
is able 
to reproduce the low value of R observed for this object although the large 
errors on the measurement of R prevent us from any firm conclusion 
(Fig.~\ref{ratio_metallicity}. In any case, 
it is likely that  
its SED gives no much room for a significant 
extinction. However this galaxy exhibits a rather high 
extinction estimated from its slope $\beta$ (equal to $-1.2$): $\rm A(1600 \AA) 
= 1.96~mag$. The extinction is more moderate but still significant if we 
use the    $\rm 
F_{FIR}/F_{UV}$ ratio and again the calibration of Calzetti et al. (see below) 
with $\rm A(1600 \AA) = ~0.8 ~mag$. The inconsistency between the  evidence for 
dust and the blue spectrum  appears in Fig.~\ref{ratiofirfuvbeta} where 
Tol1247-232 does not follow the general trends. \\

 If we exclude the case of Tol1247-232, we reach the conclusion that 
the flat FUV SED of  starburst galaxies  
are  consistent with recent starburst or continuous star formation. The 
spectra which exhibit a decrease shortward of 1050 $\rm \AA$ cannot be 
satisfactorily 
fitted by evolutionary models and require attenuation by dust for the presence
 of which we  
 already have some evidence. 
In short, the differences between the relative spectral energy distributions
 of our sample  would be caused primarily by dust extinction.

\section{The internal dust extinction}

\begin{figure}
  \includegraphics[angle=-90,width=8cm]{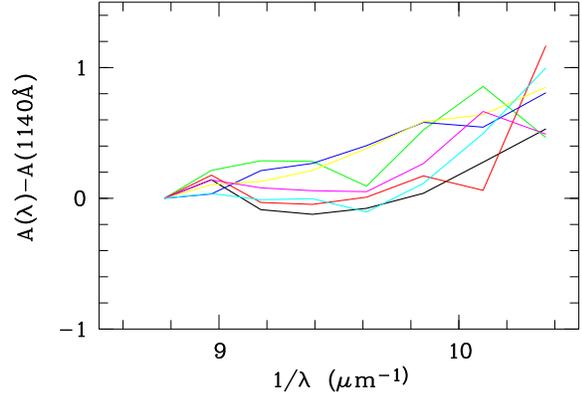}
  
     \caption{The differential extinction $\rm A(\lambda)-A (1140\AA)$ as a 
function of $\rm 1/\lambda$ in $\rm \mu m^{-1}$ for NGC3504, NGC7673, 
NGC7714, NGC3690, NGC1140, NGC3310 and NGC5236. The colors used are the same as 
in Fig.~\ref{sedcompil} }
      \label{Alambda}
         \end{figure}
         
 \begin{figure}
  \includegraphics[angle=-90,width=8cm]{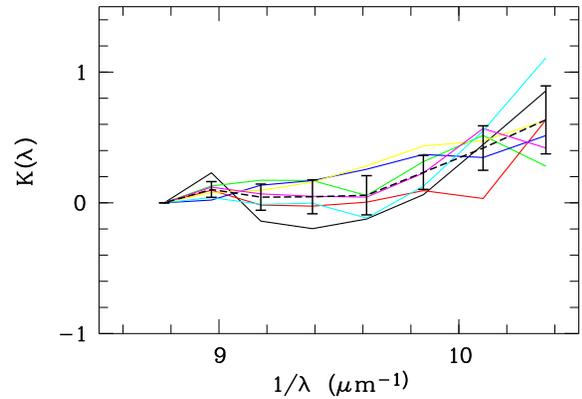}
  
     \caption{The attenuation laws normalized at 1140 $\rm \AA$ as a function of  
$\rm 1/\lambda$ in $\rm \mu m^{-1}$ for the same galaxies as in 
Fig.~\ref{Alambda}. The mean curve is drawn as a dashed line and the error bars 
indicate the standard deviation.}
      \label{klambda}
         \end{figure}        
         
The obvious presence of dust in some of our targets allows us to study the 
properties 
of dust extinction in starburst galaxies in the far-UV (900 - 1200 \AA).  
 If our conclusion that the variations of spectral shapes are essentially due to 
  dust extinction is correct, 
 we can try to derive the attenuation law for starburst  
 galaxies by 
comparing the spectra of dusty galaxies with a reference spectrum supposed to be 
without extinction. The method is similar to that used for the derivation of a 
stellar extinction law. Nevertheless in case of galaxies the 
attenuation curve  will be the result of the combined effects of dust properties 
and geometry (e.g. Calzetti et al. \cite{calzetti94}, Leitherer et al. 
\cite{leitherer02}).\\
We 
exclude from the analysis Tol1247-232 which has a spectrum bluer than any 
prediction of synthesis models. All the  spectra are corrected for the 
Galactic extinction. The spectrum of each galaxy  is rebinned according to the 
rest-frame wavelength from 965 $\rm \AA$ to 1165 $\rm \AA$ by bin of 25 $\rm 
\AA$. At this 
step, we discard MRK54 whose  data only cover the rest frame spectral range 
from 1010 to 1100 $\rm \AA$.

In order to derive the attenuation curve we  make the following steps
\begin{enumerate}
\item We normalize all the observed spectra at 1140 $\rm \AA$

\item We construct a reference spectrum  without dust by adding the spectra 
of the 3 galaxies which exhibit a flat SED namely NGC1705, Tol1924-416 and 
MRK59.  

\item We assume that the departure of the other spectra from our reference 
spectrum is due to  dust extinction alone and we deduce $\rm 
A(\lambda)-A(1140  
\AA)$ for each galaxy. 

\end{enumerate}

\begin{table}
\caption[]{The dust-free reference SED normalized at 1140 $\rm \AA$, obtained by 
adding the spectra of NGC1705, MRK59 and Tol1924-416}

\begin{flushleft}
\begin{tabular}{llllllll} 

\hline 

965 &990 &1015 &1040 &1065 &1090 &1115 &1140 \\ 
\hline 
 & & & & & & & \\
1.14    &1.11&1.07 &1.15 &1.12 & 1.07 &  1.08   &1.0\\

\hline 
\end{tabular} 
\end{flushleft} 
\end{table}

The dust-free reference spectrum is given in Table 5.
The resulting attenuation curves $\rm A(\lambda)-A(1140 \AA)$
 for each galaxy are presented in Fig.~\ref{Alambda}. The normalization at 1140 
$\rm \AA$ is motivated by the lower dispersion found for the attenuation curve 
as 
compared to a normalization at 1165 $\rm \AA$. Moreover, the  flux at 1165 $\rm 
\AA$ 
(rest-frame) corresponds to an observed wavelength range where {\it FUSE} begins 
to be 
less sensitive.

Our approach is almost similar to that of Leitherer et al. ({\cite{leitherer02}) 
but more empirical: they compared the  observed spectra to model templates for   
deriving the extinction curve whereas we prefer to construct empirically our 
reference spectrum with the galaxies exhibiting almost no extinction. Such an 
empirical approach was also adopted by Calzetti et al. (\cite{calzetti94}) to 
derive the extinction curve longward of 1200$\rm \AA$.

In order to derive an attenuation law we need to normalize $\rm 
A(\lambda)-A(1140 
\AA)$ by the amount of extinction for each target (equivalent to the color 
excess E(B-V) used to normalize  a stellar extinction law). \\
Quantifying the extinction in galaxies is a very difficult task.  The slope 
$\beta$ 
of the UV continuum longward of 1200 $\rm \AA$ and the ratio  $\rm 
F_{FIR}/F_{UV}$ 
are 
expected 
to be quantitative tracers of the amount of extinction in starburst galaxies. 
Since the $\rm F_{FIR}/F_{UV}$ ratio is only available for three targets 
(excluding 
the targets used to construct the reference spectrum) we adopt the slope $\beta$ 
to normalize the dust attenuation law. Practically we follow the same method as 
Leitherer et al. (\cite{leitherer02}).

For each reddened galaxy we calculate:
$$ \rm k(\lambda) = {{A(\lambda)-A(1140 \AA)}\over \Delta \beta}$$
where $\Delta \beta$ is the difference between the observed slope $\beta$ of the 
galaxy and the intrinsic slope of the reference spectrum $\beta_0 = -2.4$ (mean 
value for the three galaxies used to construct the reference spectrum).

The resulting attenuation laws normalized at 1140 $\rm \AA$ are reported in 
Fig.~\ref{klambda}. The average of the 7 curves gives the mean attenuation law 
$<k(\lambda)>$ and the standard deviation is taken as a measure of the error. 
The normalization only slightly decreases the 
dispersion between the curves which probably implies that the amount of 
extinction is not very accurately measured by $\Delta \beta$. Note however that 
our sample does not exhibit a large range of extinction in terms of $\rm \Delta 
\beta$.  In 
the absence of an other method to estimate the amount of extinction we go on 
with this normalization.

A polynomial fit of $<k(\lambda)>$ gives:
$$<k(\lambda)> = 32.356-7.1023/\lambda+0.38995/\lambda^2$$
with $\lambda$ expressed in $\mu$m.

In order to perform a comparison with other attenuation curves  we also derive a 
more classical 
formula of the attenuation law related to the stellar color excess $\rm 
E(B-V)_s$. To this aim 
we need 
to relate $\Delta \beta$ and $\rm A(1140 \AA)$ to $\rm E(B-V)_s$. We 
use the laws drawn by  Calzetti et al. (\cite{calzetti00}) for starburst 
galaxies,  extrapolated from 1200 to 1140 $\rm \AA$:  
$$\rm \Delta \beta = 4.72~E(B-V)_s$$ 
and 
$$\rm A(1140 \AA) = 12.75~E(B-V)_s$$
and deduce
$$\rm A(\lambda) =((4.72~<k(\lambda)>+12.75)\times E(B-V)_s$$
$$\rm =(165.470-33.5229/\lambda+1.84056/\lambda^2)~ 
E(B-V)_s$$

for the extinction of starburst galaxies between 0.095 and 0.1140$\mu$m  
($\lambda$ is expressed in $\mu$m in the formula).

\section{Discussion}

The attenuation curve obtained in the previous section is a mean curve and the 
estimated errors are rather large. As already discussed in the previous 
sections the far-UV SED of a star-forming 
galaxy is the result of the combined effects of  star formation history,  
initial mass function and   extinction. To deduce the 
attenuation curve we assumed that the departure of the observed SED from a 
template is entirely due to the internal dust extinction. The SED used as a 
dust-free 
 template is consistent with a constant star formation rate over some Myrs.
  
 We may try to test the validity of our  approach a posteriori by  applying our 
mean 
attenuation  curve   to the entire set of obscured spectra (the 7 spectra used 
to construct the attenuation curve and presented in Fig.~\ref{Alambda}) and 
perform again the 
analysis of section 4 (points 3 to 5). All the corrected spectra exhibit a ratio 
R between 1 and 1.2 except NGC3504 for which R is found equal to 0.8. Such 
values of R are consistent  with a constant star formation or a very recent 
starburst (Fig.~\ref{ratio_metallicity}).  
 We can also perform again the fit with  synthesis models   by a $\chi^2$ 
minimization. However, one must note that the attenuation curve holds only in 
the 
range 950 to 1140 $\rm \AA$ and leads to only 6 to 11 values per spectrum. In 
all cases the fitting procedure 
selects models with a recent star formation consistent with our dust-free 
template. The  
$\chi^2$ values are lower than 1.5 for 5  of the 7 galaxies analysed. The two 
discrepant cases are NGC3504 as expected from the value of its ratio R (0.8) and 
NGC5236 (but for 
this object we have only 6 values for the SED). By comparison, the same 
procedure 
 performed in section 4 on the  observed spectra of  these 7 galaxies selected  
old starbursts with larger $\chi^2$ values ($>2$) and did not reproduce 
correctly 
the values of R. 
 
Therefore, the internal dust extinction is likely to be the dominant factor of
the departure of the SEDs from the dust-free template.  Nevertheless, whereas
systematic effects are unlikely, we cannot exclude that the star formation
history within each object has some (secondary) influence on the SED.  This
detailed star formation history combined with  measurement errors and the
uncertainty on the normalization of the attenuation law (cf.  section 5) can
easily explain the dispersion found around  the mean attenuation curve.\\

We can compare the attenuation curve  derived with other curves measuring the 
extinction in the far ultraviolet. Very few are available. Leitherer et al. 
({\cite{leitherer02}) deduced a mean attenuation curve from 900 to 1800 $\rm 
\AA$ 
for 
ten star forming galaxies with a method similar to ours. Because of its  
popularity we  also 
extrapolated the law proposed by Calzetti et al. (\cite{calzetti00}) for 
starburst galaxies down to 965 $\rm \AA$ (10.34 $\rm \mu m^{-1}$)  
 even if it  was not derived for this wavelength range. We also 
compare our attenuation 
curve to the stellar extinction curve obtained by Sasseen et al.  
(\cite{sasseen02}). The comparison between these curves is illustrated in 
Fig.~\ref{comp_ext}.

\begin{figure}
  \includegraphics[angle=-90,width=8cm]{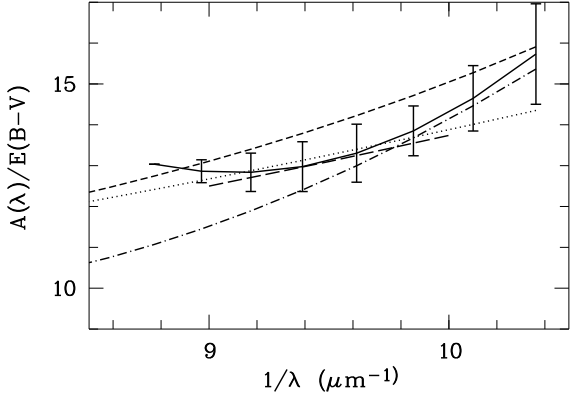}
  
     \caption{Comparison between  curves measuring the extinction in the far 
ultraviolet for galaxies and stars. The solid lines and the error bars refer to 
the attenuation 
curve derived in this paper. The dotted line is the attenuation curve of 
Leitherer et al. (\cite{leitherer02}); the short dashed line is the 
extrapolation of 
the attenuation law of Calzetti et al. (\cite{calzetti00}). The dot-dashed line 
is the stellar extinction curve of Sasseen et al. (\cite{sasseen02}). The long 
dashed line is the model of Witt and Gordon (\cite{witt00}) for a shell 
distribution, a clumpy dust and an optical depth in the V Band equal to 1.5}
      \label{comp_ext}
         \end{figure}
The attenuation law derived in the present work is found almost flat from 8.8 to 
9.6 $\rm \mu m^{-1}$ (1140 
to 1040 $\rm \AA$) and rises at lower wavelengths. 
Given the error bars (only ours are reported) the various curves presented in 
Fig.~\ref{comp_ext} are consistent. Nevertheless, applying the attenuation law 
derived in this present work leads to a larger extinction than using that of 
 Leitherer et al. (\cite{leitherer02}) for $\rm \lambda < \leq 1050 \AA$. The 
difference in 
$A(\lambda)/E(B-V)$ reaches 1.4 units at 965 $\rm \AA$. In this wavelength range 
($\rm \lambda < \leq 1050 \AA$) our curve is very similar to the stellar 
extinction 
curve of Sasseen et al. (\cite{sasseen02}). Note that the extrapolation of the 
Calzetti et al. (\cite{calzetti00}) 
attenuation law gives a larger extinction  than the other laws.
  
 Witt \& Gordon (\cite{witt00}) proposed several models of dust extinction 
in galaxies. The adopted geometry is spherical with a clumpy, dusty or shell 
distribution of the stars and the dust. In particular, they  were able to 
reproduce the attenuation law of Calzetti (\cite{calzetti97}) assuming a shell 
distribution, a 
clumpy dust medium and a mean optical depth in the V band equal to 1.5. Their 
predictions extend down to 1000 $\rm \AA$ and we report them in 
Fig.~\ref{comp_ext}. The model predicts a greyer extinction than our attenuation 
law 
 very consistent with the attenuation law of Leitherer et al. The model appears 
below the extrapolation of the law of Calzetti at al. (\cite{calzetti00}); this 
is likely to be due to the re-calibration of the attenuation law between 1997 
and 2000,  the model being adjusted on an older version of the 
attenuation law  (Calzetti  \cite{calzetti97}).

\section{Conclusion}

We  obtained the   spectral energy distribution of 
12 nearby
starburst 
galaxies from 970 to 1175 $\rm \AA$. The high spectral resolution of {\it 
FUSE}  allowed us to properly estimate the continuum out of the interstellar 
lines. The general behavior is a rather flat 
distribution, in general agreement with the predictions of population synthesis 
models in 
this wavelength range for an active star formation. 
 Nevertheless a  decrease of the flux downward 1050 $\rm \AA$ is observed in   
some 
 objects: the mean 
ratio of the fluxes at 1070 and 1010 $\rm \AA$ is found to correlate with the 
metallicity and the extinction of the galaxies traced by their $\rm 
F_{FIR}/F_{UV}$ ratio or the slope $\beta$ of the spectral energy distribution 
longward of 1200 $\rm \AA$.\\
The galaxies   with an 
almost flat spectrum   have almost no extinction and a sub-solar metallicity 
(0.1 to 0.2$\rm 
Z_\odot$);   
 their spectrum is well fitted by  a very 
young instantaneous burst of star formation or 
a constant  star formation and a Salpeter IMF. The FUV spectra of 
the galaxies which exhibit a decrease 
of the flux downward $\rm 1050 \AA$ cannot be accurately fitted by any of the 
models 
we consider. These galaxies show evidence for dust extinction and have in 
general a metallicity from $\rm  0.4 
Z_\odot$ to  
 $\rm 2 Z_\odot$. 
\\

We deduce a relative dust  attenuation law in the range 965-1140 $\rm \AA$ by 
comparing the SED of the sample galaxies with a dust free template constructed 
with the SED of the galaxies which do not exhibit any trace of extinction. The 
total amount of extinction is measured with the observed slope of the UV 
continuum longward of 1200 $\rm \AA$. We also derive an attenuation law related 
to 
the 
stellar color excess $A(\lambda)/E(B-V)_s$ by using (and sligtly extrapolating) 
the calibrations of Calzetti et al. (\cite{calzetti00}). 
We give a simple polynomial parametrization of $A(\lambda)/E(B-V)_s$.
The extinction that we find is almost constant from 1165 to 1040 $\rm \AA$ and 
rises 
at  
lower wavelengths. The agreement with the attenuation law proposed by Leitherer 
et al. (\cite{leitherer02}) is good at $\rm \lambda > 1040 \AA$ but we find a 
larger  extinction than 
Leitherer et al. at $\rm \lambda < 1040 \AA$. Nevertheless, given the error bars 
in  
both studies,  the attenuation laws remain consistent. 

\begin{acknowledgement}
	This research is based on observations made with the NASA-CNES-CSA Far
Ultraviolet Spectroscopic Explorer.  FUSE is operated for NASA by the Johns
Hopkins University under NASA contract NAS5-32985.\\ 
This research has also made use
of the NASA/IPAC Extragalactic Database (NED) which is operated by the Jet
Propulsion Laboratory, California Institute of Technology, under contract with
the National Aeronautics and Space Administration.
\end{acknowledgement}


\begin{thebibliography}{}
\bibitem[2000]{barnstedt00} Barnstedt, J., Gringel, W., Kappelmann, N., Grewing, 
M. 2000, A\&AS 143, 193
\bibitem[2001]{bell01} Bell, E.F., Kennicutt, R.C. 2001, ApJ 548, 681
\bibitem[1989]{cardelli89} Cardelli, J. A., Clayton G. C., Mathis, J. S. 1989, 
ApJ 430, 630
\bibitem[1994]{calzetti94} Calzetti, D., Kinney A.L., Storchi-Bergmann T. 1994, 
ApJ 429, 582
\bibitem[1997]{calzetti97} Calzetti, D. 1997, AJ 113, 162
\bibitem[2000]{calzetti00} Calzetti D., Armus, L., Bohlin, R.C., Kinney, A.L.,      
Koornneef, J., Storchi-Bergmann T. 2000, ApJ 533, 682
\bibitem[1994]{deharveng94} Deharveng, J.M., Sasseen T.P., Buat, V., Bowyer, S., 
Wu, X. 1994, A\&A 289, 71
\bibitem[1987]{donas87} Donas, J, Deharveng, J.M., Milliard, B., Laget, M., 
Huguenin, D. 1987,  A\&A 180, 12 
\bibitem[1997]{gonzalez97} Gonzalez Delgado, R., Leitherer, C., Heckman, T. 
1997, ApJ, 489, 601 
\bibitem[1998]{heckman98} Heckman, T.M., Robert, C., Leitherer, C., Garnett, 
D.R., 
van der Rydt, F. 1998, ApJ 503, 646  
\bibitem[2001a]{heckman01a} Heckman, T.M., Sembach, K.R., Meurer, G.R., 
Strickland, D.K., 
Martin, C.L., Calzetti, D., Leitherer, C. 2001 ApJ 554, 1021
\bibitem[2001b]{heckman01b} Heckman, T.M., Sembach, K.R., Meurer, 
G.R.,Leitherer, C., Calzetti, D., Martin, C.L. 2001 ApJ 558, 56
 \bibitem[1988]{helou88} Helou, G., Khan, I.R., Malek, L., Boehmer, L. 1988, 
ApJSS 
68, 151 
\bibitem[1996]{giavalisco96} Giavalisco, M., Koratkar, A., Calzetti, D. 1996, 
ApJ, 466, 831
\bibitem[1993]{kinney93} Kinney, A.L., Bohlin, R.C., Calzetti, D., Panagia, N., 
Wyse, 
R.F. 1993, ApJSS 86, 5
\bibitem[1995]{leitherer95} Leitherer, C.,  Heckman, T.M. 1995, ApJS 96, 9
\bibitem[1999]{leitherer99} Leitherer, C., Schaerer, D., Goldader, J.D. et al. 
1999, ApJS 123, 3
\bibitem[2002]{leitherer02} Leitherer, C., Li, I-H, Calzetti, D., Heckman, T.M.  
2002 ApJSS submitted
\bibitem[1995]{meurer95} Meurer G.R., Heckman T.M., Leitherer C., Kinney 
A.L., Robert C., Garnett D.R., 1995, AJ 110, 2665 
\bibitem[1999]{meurer99} Meurer G.R., Heckman T.M., Calzetti D., 1999, 
ApJ 521, 64
\bibitem[1994]{milliard94} Milliard, B., Donas, J., Laget, M., Huguenin, D. 
1994, {\it The balloon-borne 40-cm UV imaging telescope FOCA; results and 
perspective}, 11th ESA Symposium, Montreux
\bibitem[2000]{noeske00} Noeske, K.G., Guseva, N.G., Fricke, K.J., Izotov, Y.I., 
Papaderos, P., Thuan, T.X.  2000, A\&A 361, 33
\bibitem[2002]{sasseen02} Sasseen, T.P., Hurwitz, M., Dixon, W.V., Airieau, S. 
 2002, ApJ 566, 267
\bibitem[1998]{schlegel98} Schlegel, D.J., Finkbeiner, D.P., Davis, M. 1998, ApJ 
500, 525 
\bibitem[2000]{shull00} Shull, J.M., Tumlinson, J., Jenkins, E.B. et al. 2000,  
ApJ 538, 
L73
\bibitem[1997]{stecher97} Stecher, T.P., Cornett, R.H., Greason, M.R., Landsman, 
W.B., Hill, J.K., Hill, R.S. et al. 1997, PASP, 109, 584
\bibitem[1993]{terlevich93} Terlevich, E., Diaz, A., Terlevich, R., Vargas, M. 
1993, MNRAS 260, 3 
\bibitem[2002]{thuan02} Thuan, T., Lecavelier des Etangs, A., Izotov, Y.I.  
 2002, ApJ 565, 941
\bibitem[2000]{vidal00} Vidal-Madjar, A., Kunth, D., Lecavelier des Etangs, A. 
et al. 2000, ApJ 538, L77 
\bibitem[2000]{witt00} Witt A. N., Gordon K. D. 2000, ApJ, 528, 799
 
\end{thebibliography}
\end{document}